\documentclass[conference]{IEEEtran}
\IEEEoverridecommandlockouts
\usepackage{cite}
\usepackage{url}
\usepackage{amsmath,amssymb,amsfonts}
\usepackage{algorithmic}
\usepackage{graphicx}
\usepackage{textcomp}
\usepackage{xcolor}
\usepackage{subfiles}
\usepackage{listings}
\usepackage{enumitem}
\usepackage{subcaption}
\usepackage{tabularx} % Required for the tabularx environment
\usepackage{booktabs} % For professional-looking tables
\usepackage{ragged2e} % For better text alignment in columns
\usepackage{float}
\usepackage{algorithm}
\def\BibTeX{{\rm B\kern-.05em{\sc i\kern-.025em b}\kern-.08em T\kern-.1667em\lower.7ex\hbox{E}\kern-.125emX}}
\begin{document}

\title{NimbusGuard: A Novel Framework for Proactive Kubernetes Autoscaling Using Deep Q-Networks}

\author{\IEEEauthorblockN{Chamath Wanigasooriya}
\IEEEauthorblockA{\textit{Department of Computer Science } \\
\textit{Informatics Institute of Technology}\\
Sri Lanka \\
chamath.2023@iit.ac.lk}
\and
\IEEEauthorblockN{Indrajith Ekanayake}
\IEEEauthorblockA{\textit{Department of Computer Science} \\
\textit{Informatics Institute of Technology}\\
Sri Lanka \\
indrajith.e@iit.ac.lk }
}
\maketitle
\begin{abstract}
Cloud native architecture is about building and running scalable microservice applications to take full advantage of the cloud environments. Managed Kubernetes is the powerhouse orchestrating cloud native applications with elastic scaling. However, traditional Kubernetes autoscalers are reactive, meaning the scaling controllers adjust resources only after they detect demand within the cluster and do not incorporate any predictive measures. This can lead to either over-provisioning and increased costs or under-provisioning and performance degradation. We propose NimbusGuard, an open-source, Kubernetes-based autoscaling system that leverages a deep reinforcement learning agent to provide proactive autoscaling. The agent’s perception is augmented by a Long Short-Term Memory model that forecasts future workload patterns. The evaluations were conducted by comparing NimbusGuard against the built-in scaling controllers, such as Horizontal Pod Autoscaler, and the event-driven autoscaler KEDA. The experimental results demonstrate how NimbusGuard's proactive framework translates into superior performance and cost efficiency compared to existing reactive methods.

\end{abstract}

\begin{IEEEkeywords}
Elasticity, Autoscaling, Microservices, Reinforcement Learning, Proactive Scaling\end{IEEEkeywords}

\section{Introduction}
Popularity in microservice and container-based approaches brought the term cloud native to the light. Kratzke and Quint \cite{ind1} defined cloud native applications as distributed, elastic systems designed to take full advantage of cloud environments. These applications are composed of small, independent, and deployable units known as microservices. The elasticity is provided through dynamic resource allocation to microservices through scaling them properly on demand \cite{ind5}. Kubernetes \cite{cham10} has become the de facto standard for microservice (container) orchestration, which handles elasticity with many of the previously mentioned properties \cite{ind2}. Kubernetes has built‑in mechanisms for dynamically allocating resources to constituent containers and scaling them on demand. However, the current approaches are reactive, meaning they adjust resources only after they detect demand within the cluster. This has proven insufficient for dynamic production workloads often leading to either under-provisioning or over-provisioning of the resources \cite{ind3,ind4}. An under‑provisioned microservice deployment cannot handle workloads efficiently, whereas an over‑provisioned deployment incurs unnecessary cost. Therefore, dynamic resource allocation that is both efficient and cost‑effective remains a challenging task. 

% These reactive, threshold-based systems lag behind real-time demand, leading to performance degradation during traffic spikes and significant resource waste during idle periods. Their simplistic approach lacks the predictive foresight and contextual understanding necessary for efficient cloud-native resource management.

To address these limitations, this paper presents NimbusGuard, a novel, multi-modal framework for proactive Kubernetes autoscaling. It introduces predictive foresight and contextual understanding necessary for efficient cloud-native resource management. The proposed solution is threefold: the Deep Q‑Network (DQN) agent learns optimal scaling policies, the Long Short‑Term Memory (LSTM) network forecasts future workloads to provide temporal awareness, and the Large Language Model (LLM) cognitive agent is orchestrated as a stateful reasoning workflow, validating and refining scaling decisions. Implemented as a production-ready Kubernetes operator, our framework uses a central Model Context Protocol (MCP) server to facilitate real-time, message-driven communication between the distributed AI agents. 

\noindent  Key Contributions:
\begin{enumerate}[label=\arabic*)]
  \item A threefold DQN–LSTM–LLM architecture for intelligent, proactive autoscaling of cloud‑native applications.
  \item The first use of a LangGraph‑orchestrated \cite{cham9} LLM agent to validate infrastructure‑scaling decisions by enriching them with real‑time log and policy data.
  % \item A replicable, open‑source feature‑engineering pipeline for Kubernetes observability data.
\end{enumerate}

% The foundation of our AI models is a scientifically rigorous feature engineering pipeline. Rather than relying on default metrics, we developed an automated process that transforms over 100 raw system metrics into a minimal, yet highly predictive, feature set.

This paper is structured as follows. Section II surveys the related literature. Section III presents the proposed proactive Kubernetes autoscaling framework. Section IV explains the experimental setup, data collection, and load generation procedure. Section V analyses the results and discusses the main insights. Section VI summarises the contributions and suggests directions for future research.
\section{Related Work} 
Elasticity in cloud‑native applications is achieved through dynamic resource allocation. It accommodates end‑user‑driven fluctuations by adjusting storage, compute, and networking resources over time. An autoscaler usually decides how many resources an application receives, increasing or decreasing capacity in real time to match user demand \cite{ind5}. Kubernetes-orchestrated microservice environments have built-in autoscaling at two different levels: At the inference level cluster autoscaler (CA) manages the elasticity property in the Nodes. At the application level, Horizontal Pod Autoscaler (HPA), Vertical Pod Autoscaler (VPA), and Kubernetes Event‑Driven Autoscaler (KEDA) \cite{ind6} manage the elasticity property in the Containers. HPA scales workloads horizontally by adding or removing pod replicas in response to resource metrics, whereas the VPA resizes individual pods by adjusting their CPU and memory requests and limits. Unlike HPA and VPA, KEDA is not part of the core Kubernetes distribution, it's a Cloud Native Computing Foundation (CNCF) graduated open-source extension for Kubernetes native HPA, enabling event‑driven scale‑out and scale‑to‑zero \cite{ind7}. \\
\indent All Kubernetes built-in solutions are reactive, so autoscaling decisions are made solely from the system’s current state \cite{ind8}. Researchers have highlighted key issues of this reactive autoscaling \cite{cham1,cham2,cham3}. Toka et al. \cite{ind4} identified three main drawbacks of the current process: (1) scaling is reactive and purely observation‑based, (2) scaling behavior is not adapting to the current variability of the demand, and (3) defining scaling behavior is cumbersome because it requires tuning numerous parameters. The same study proposed a Machine Learning (ML) based proactive scaling engine. Mondal et al. \cite{cham1} highlighted the same issue of reactive autoscaling using CPU and memory as a metric, making HPA incapable of foreseeing upcoming workload spikes leading to Quality of Service (QoS) violations, long tail latency, and wasted resources. The authors proposed a proactive Custom Pod Autoscaler that uses a Gated Recurrent Unit (GRU) based load prediction model and a stability window to scale pods ahead of demand. \\
\indent To address the aforementioned issues, researchers have explored proactive autoscaling \cite{cham2,ind4,ind9} or hybrid autoscaling \cite{ind10,cham6} methods based on time series algorithms. In terms of proactive autoscaling, the literature reveals three themes of implementation: (1) Short-term demand forecasting based, (2) performance prediction based, and (3) Reinforcement Learning (RL) based. Each implementation exhibits inherent drawbacks. Predictive models, such as the GRU by Mondal et al. \cite{cham1} and the Bi-LSTM by Dang-Quang and Yoo \cite{cham2}, improve upon HPA by forecasting workloads. However, they function as open-loop systems. They predict future demand but lack a sophisticated, closed-loop mechanism to learn from the real-world impact of their scaling decisions. RL approaches from Khaleq et al. \cite{cham3} and Garí et al. \cite{cham4} introduce adaptive decision-making but are often constrained. For example, Khaleq et al. \cite{cham3} focus on learning optimal thresholds rather than selecting direct scaling actions, which is an indirect and less agile control method. Furthermore, these RL models, including the DQN-based scheduler from  Jian et al. \cite{cham5}, operate on a limited set of default metrics and act as black boxes, lacking an engineered feature set and explicit reasoning capabilities. A significant gap in all reviewed literature is the absence of a cognitive validation layer. Also, they lack a mechanism to interpret unstructured data like real-time logs or to apply stateful reasoning to validate a scaling decision before execution. This exposes them to risks where a purely quantitative model might scale inappropriately during complex events like a canary deployment or a database migration \cite{cham7}.

%\subfile{../tables/related_work}

\section{Methodology}
\subsection{Overview}\label{sec:methodology_overview}
\begin{figure}[htbp]
\centering
% The image width is set to 80% of the column's width.
\includegraphics[width=0.8\columnwidth]{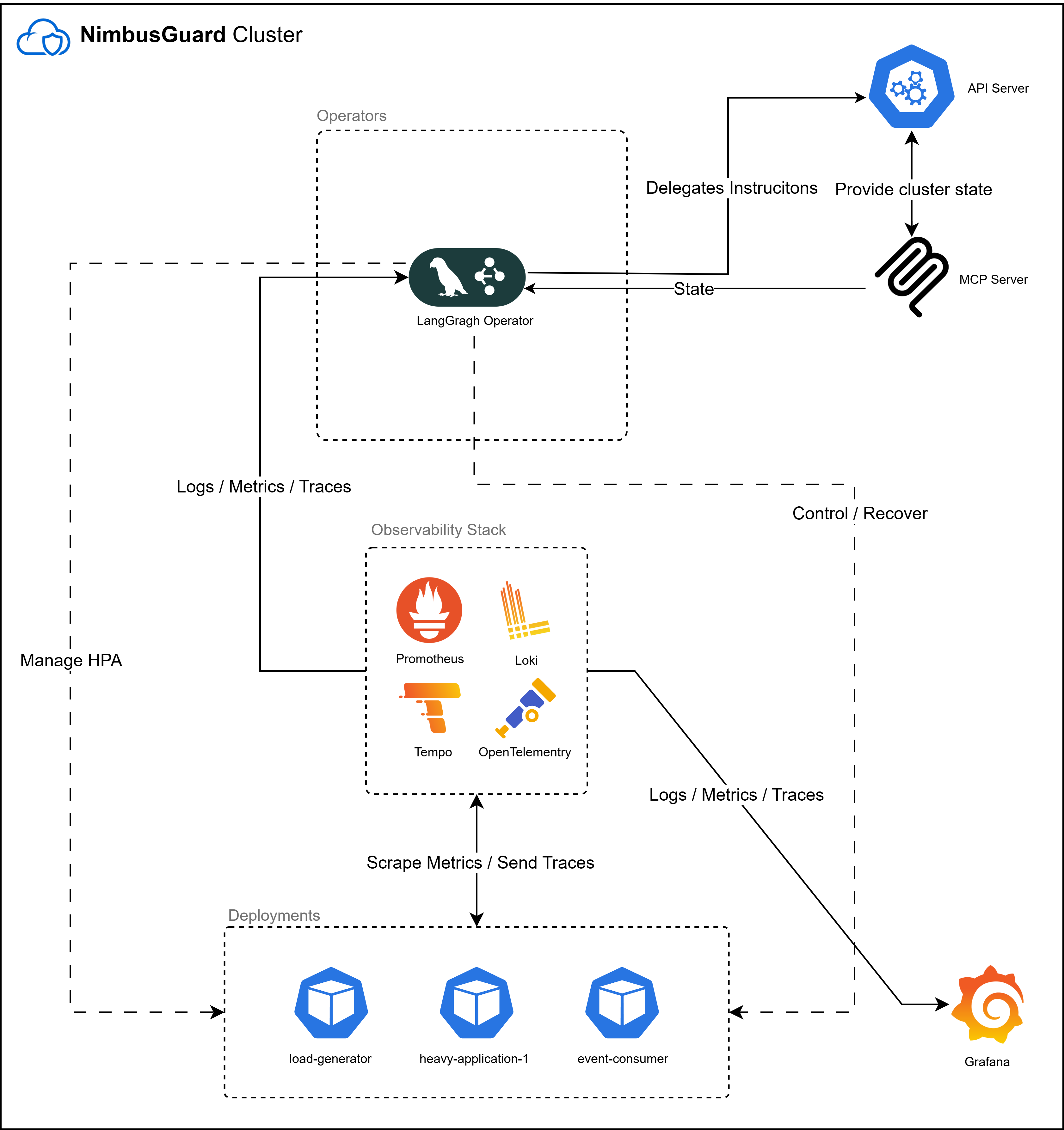}
\caption{High-level Overview of the Framework}
\label{fig:high_level}
\end{figure}

NimbusGuard framework ensures component integration between the DQN adapter, LSTM forecaster, native Kubernetes API, and Prometheus monitoring stack. The architecture, as depicted in Figure \ref{fig:high_level}, summarizes the overall flow of our proposed approach, and it can be divided into the following decision pipeline. Data Collection → Feature Processing → DQN Inference → LLM Validation → Scaling Execution. This cycle operates in 15-second intervals with stabilization periods.
% \begin{figure}[htbp]
% % Use \centering for better spacing control within the figure environment.
% \centering
% % Set the width of the image to the width of the current column.
% \includegraphics[width=\columnwidth]{../assets/architecture.png}
% \caption{High-level Overview of the Framework}
% \label{fig:high_level}
% \end{figure}

\subsection{Algorithmic Framework}
NimbusGuard implements a novel hybrid autoscaling algorithm that combines Deep Q-Network (DQN), Long Short-Term Memory (LSTM) forecasting, and an optional Large Language Model (LLM) validation layer for intelligent Kubernetes container scaling. The system operates on a 30-second decision interval. At each interval, it constructs a 6-dimensional state vector. The agent's action space is a discrete set of three actions: scale\_down (-1), keep\_same (0), and scale\_up (+1). Upon initialization, the system loads its core components:

\begin{itemize}
    \item \textbf{DQN Networks:} A Dueling DQN architecture is used for both the main and target networks to improve policy evaluation.
    \item \textbf{LSTM Forecaster:} A 2-layer LSTM with 32$\rightarrow$16 hidden units is used for time-series prediction of Kubernetes consumer pod memory usage. The model uses only 2 aggregated features $\text{total\_memory\_mb}, \text{pod\_count}$ with a 20-interval lookback window \textit{5 minutes} and achieves 8.7\% MAPE accuracy for 15-second ahead predictions.
    \item \textbf{Experience Replay Buffer:} A buffer with a capacity of 10,000 experiences is used for training the DQN agent.
    \item \textbf{Persistent Storage:} Pre-trained models and feature scalers are loaded from a MinIO object store.
\end{itemize}
The main operational loop is orchestrated as a stateful graph using LangGraph, ensuring a modular and traceable execution of the six critical nodes at each 30-second interval.

\subsection{Context-Aware Reward System}
The context-aware reward function is designed to guide the learning agent toward optimal resource management decisions by dynamically balancing performance, efficiency, and stability. The algorithm's has the ability to adjust its reward composition based on the prevailing system context, which is determined by workload characteristics and forecast confidence.

The reward calculation begins by evaluating three components. The first is a performance score that quantifies the quality of service (QoS) using metrics. The second is an efficiency score that assesses resource utilization, rewarding configurations that meet performance targets with fewer resources. The third component is a stability penalty, introduced to discourage volatile or excessively frequent scaling actions and promote system stability.

 The system's current operational state is classified by its workload level (e.g., low, nominal, or high load). This classification, along with the confidence of the workload forecast, determines two critical weights: one for the current state and one for the forecasted state. This allows the reward system to prioritize different objectives under varying conditions. For instance, during a high-load state with a high-confidence forecast, more emphasis can be placed on the proactive, forecast-based reward component.

% The total reward is a composite of two primary terms: the `{$current\_reward$}` and the `{$forecast\_reward$}`.

\begin{enumerate}

    \item \textbf{Current State Utilization Reward ($R_{\text{current}}$)}:
    This evaluates the immediate system performance using Gaussian reward curves centered on optimal resource utilization targets. It combines CPU utilization reward (target: 70\%) and memory utilization reward (target: 80\%) with deployment-specific normalization:
    $$R_{\text{current}} = w_{\text{cpu}} \cdot R_{\text{cpu}}(u_{\text{cpu}}) + w_{\text{mem}} \cdot R_{\text{mem}}(u_{\text{mem}})$$

    \item \textbf{Forecast-based Proactive Reward ($R_{\text{forecast}}$)}:
    When LSTM-based memory predictions are available, this encourages proactive scaling decisions by evaluating the forecasted system state. The forecast reward is calculated using the same utilization reward function but applied to predicted metrics:
    $$R_{\text{forecast}} = R_{\text{utilization}}(\text{predicted\_metrics})$$

    \item \textbf{Combined Utilization Reward}:
    The system weights current and forecast rewards with emphasis on predictive capabilities:
    $$R_{\text{combined}} = w_{\text{current}} \cdot R_{\text{current}} + w_{\text{forecast}} \cdot R_{\text{forecast}}$$

    \item \textbf{Stability and Cost Components}:
    Additional components include stability rewards for maintaining system health, action-specific bonuses for appropriate scaling decisions, and cost penalties for resource waste. The final reward integrates all components:
    $$R_{\text{total}} = R_{\text{combined}} + R_{\text{stability}} + R_{\text{action\_bonus}} - R_{\text{cost\_penalty}}$$
    The system includes action-specific bonuses for appropriate scaling (+0.2 for scale-up, +0.15 for scale-down) and penalties for unnecessary actions ($-$0.3 for unnecessary scaling, $-$0.5 for thrashing behavior).    
\end{enumerate}

\subsection{LangGraph Stateful Ochestration}
Algorithm \ref{alg:langgraph_workflow} presents an autonomous scaling workflow implemented as a directed acyclic graph (DAG). The \textbf{ExecuteWorkflow} function initializes system state $\mathbf{S}$ with current replica count $r_c$ and processes it through six sequential nodes:

\begin{enumerate}
    \item \textbf{CollectMetrics}: Gathers real-time performance data
    \item \textbf{GenerateForecast}: LSTM network predicts future resource demands
    \item \textbf{DQNDecision}: DQN agent determines optimal scaling action
    \item \textbf{ValidateDecision}: MCP server enforces replica limits, scaling velocity constraints, and mandatory cool-down periods to prevent system thrashing
    \item \textbf{ExecuteScaling}: Implements the validated scaling decision
    \item \textbf{CalculateReward}: Computes reward signal for DQN policy refinement
\end{enumerate}

The workflow ensures system stability through MCP validation checks while enabling continuous learning via reward feedback. The function returns updated system state $\mathbf{S}$, completing the adaptive scaling cycle.
% Algorithm 2: LangGraph Workflow
\begin{algorithm}
\caption{LangGraph Workflow Execution}
\label{alg:langgraph_workflow}
\begin{algorithmic}[1]
\REQUIRE Current replica count $r_c$
\ENSURE Updated system state $\mathbf{S}$
\STATE $\mathbf{S} \leftarrow \text{InitializeState}(r_c)$
\STATE $\mathbf{S} \leftarrow \text{CollectMetrics}(\mathbf{S})$ \COMMENT{Node 1: Metrics Collection}
\STATE $\mathbf{S} \leftarrow \text{GenerateForecast}(\mathbf{S})$ \COMMENT{Node 2: LSTM Forecasting}
\STATE $\mathbf{S} \leftarrow \text{DQNDecision}(\mathbf{S})$ \COMMENT{Node 3: DQN Decision Making}
\STATE $\mathbf{S} \leftarrow \text{ValidateDecision}(\mathbf{S})$ \COMMENT{Node 4: Safety Validation}
\STATE $\mathbf{S} \leftarrow \text{ExecuteScaling}(\mathbf{S})$ \COMMENT{Node 5: Scaling Execution}
\STATE $\mathbf{S} \leftarrow \text{CalculateReward}(\mathbf{S})$ \COMMENT{Node 6: Reward \& Learning}
\RETURN $\mathbf{S}$
\end{algorithmic}
\end{algorithm}

\subsection{Feature Engineering and Selection}\label{sec:feature_engineering}
To optimize the learning efficiency of the DQN agent, a feature engineering process was undertaken to reduce the state space dimensionality. The primary goal was to create a state vector that is both computationally efficient and information-rich, eliminating redundancy while retaining the most critical signals for intelligent autoscaling. This resulted in a compact 4-dimensional state vector, which distills complex system metrics into a focused representation of the deployment's current state and predicted resource needs.

The final state vector, $S$, is defined as: 
$$S = [s_1, s_2, s_3, s_4]$$

The components are constructed dynamically using real-time metrics and predictive modeling:

\begin{itemize}
\item[\textbf{$s_1$:}] \textbf{Predicted Memory Utilization (\%).} The forecasted memory utilization percentage based on LSTM time-series prediction, providing proactive insight into future memory pressure. This is the primary predictive signal for scaling decisions.

\item[\textbf{$s_2$:}] \textbf{Current CPU Utilization (\%).} The instantaneous CPU utilization percentage relative to the deployment's total CPU limits, calculated as: 
$$s_2 = \frac{\text{Total Current CPU Usage}}{\text{Total CPU Limit}} \times 100$$

\item[\textbf{$s_3$:}] \textbf{Current Memory Utilization (\%).} The instantaneous memory utilization percentage relative to the deployment's total memory limits, calculated as:
$$s_3 = \frac{\text{Total Current Memory Usage}}{\text{Total Memory Limit}} \times 100$$

\item[\textbf{$s_4$:}] \textbf{Current Replica Count.} The absolute number of currently active replicas, providing direct awareness of the current scaling state and serves as a baseline for scaling actions.
\end{itemize}

This approach for feature engineering ensures that the agent operates on the most essential signals while maintaining the ability to make informed, proactive scaling decisions. The focus on memory prediction as the primary forward-looking signal reflects the critical importance of memory management in containerized environments, where memory pressure can lead to pod evictions and service degradation.

\section{Experimental Setup}
This section details the environment, application, and methodologies used to conduct a comparative analysis of the three autoscaling configurations.

\subsection{Testbed Environment}
All experiments were conducted on a MacBook Pro equipped with an Apple M4 Pro processor and 24GB of unified memory. The testbed utilized Docker Desktop for Mac (v4.x), which provided the containerization runtime. The Kubernetes environment was a KinD (Kubernetes-in-Docker) cluster running Kubernetes v1.28, provisioned and managed by Docker Desktop. A significant portion of the host machine's resources (specifically, 8 vCPU cores and 16GB of memory) were allocated to the Docker Desktop virtual machine to ensure the KinD cluster had sufficient and stable resources for the experiment. The entire test was executed within a dedicated Kubernetes namespace to ensure workload isolation.

\subsection{Target Application}
The workload consisted of a stateless, containerized application developed in Python with the FastAPI framework. The application was designed as a deterministic consumer, where each incoming request triggers a predictable and consistent amount of CPU and memory usage. This deterministic behavior makes it an ideal testbed for evaluating and comparing the responses of different autoscaling mechanisms. Each container replica was configured with the following resource specifications:
\begin{itemize}
    \item CPU Request: 600m (0.6 of a virtual core)
    \item CPU Limit: 1000m (one virtual core)
    \item Memory Request: 512Mi
    \item Memory Limit: 1Gi
\end{itemize}
This configuration ensures that CPU utilization is the primary scaling signal, providing a clear metric for the autoscalers to act upon.

\subsection{Autoscaling Configurations}
Three autoscaling configurations were evaluated, representing proactive, reactive, and event-driven paradigms:

\begin{table}[htbp]
\centering
\caption{Comparison of the three autoscaling configurations evaluated.}
\label{tab:autoscaling_configs}
\begin{tabularx}{\columnwidth}{l l X}
\toprule
\textbf{Autoscaler} & \textbf{Paradigm} & \textbf{Key Configuration Details} \\
\midrule
\textbf{NimbusGuard} & Proactive & Uses a DQN agent with LSTM memory forecasting and a 4-dimensional state vector. \\
\midrule
\textbf{HPA} & Reactive & A standard Kubernetes baseline triggered by 70\% CPU or 80\% memory usage, with a 30-second stabilization window. \\
\midrule
\textbf{KEDA} & Event-driven & Configured to use Prometheus metrics with a 30-second polling interval and cooldown period to match HPA. \\
\bottomrule
\end{tabularx}
\end{table}

\subsection{Load Generation and Procedure}
We employed a fire-and-forget asynchronous load testing methodology \cite{ind11} to simulate realistic, unconstrained traffic. This approach prevents the load generator from becoming a bottleneck and allows the system's true performance under pressure to be observed.

A custom Python script utilizing `asyncio` \cite{cham8} was used to generate the load. To ensure deterministic and perfectly reproducible traffic patterns for fair comparison across all three autoscalers, the load generation process was initialized with a fixed seed.

\noindent The experiment followed a phased procedure:
\begin{itemize}
\item Phase 1: Ramp-up: A gradual increase in load (4 concurrent users, 40 total requests) to test the initial responsiveness of each autoscaler.
\item Phase 2: Sustained Load: A period of consistent, high load (8 concurrent users, 60 total requests) to evaluate steady-state behavior and stability.
\item Phase 3: Peak Load: A significant spike in traffic (15 concurrent users, 90 total requests) to challenge the system's maximum scaling capabilities and stress resilience.
\item Phase 4: Cooldown: A cessation of traffic with a light load (3 concurrent users, 30 total requests) to observe scale-down behavior and resource de-allocation efficiency.
\end{itemize}

\subsection{Data Collection and Metrics}
System-wide metrics were collected using a Prometheus monitoring instance deployed within the cluster, configured with a scrape interval of 15 seconds for high-resolution data. The primary metric analyzed for this study was the number of active application replicas over the duration of the experiment. This metric reflects the scaling decisions made by each controller in response to the identical, reproducible load pattern.
\section{Results}
\subsection{Performance Metrics Comparison}\label{sec:results_metrics}

% \subfile{../figures/hpa_baseline}
% \subfile{../figures/keda_baseline}
% \subfile{../figures/dqn_baseline}
The results obtained by subjecting each autoscaler to an identical, phased load pattern (discussed in the load generation section) reveal a clear divergence between the different systems. The DQN-based NimbusGuard demonstrated a highly aggressive, performance-oriented strategy, while HPA and KEDA exhibited a more conservative, resource-efficient approach. This aggressive strategy is a direct consequence of the hyperparameters chosen for the DQN agent, which were tuned to prioritize future performance. By assigning a heavy weight to forecasted metrics \(forecastweight=0.7\). Specifically, a high discount factor \(gamma=0.99\) makes the agent farsighted, encouraging it to scale up proactively now to prevent future negative rewards associated with performance degradation. This forward-looking behavior is coupled with a setup designed for agility; a relatively high learning rate \(lr=0.001\) and a small replay buffer \(buffer=10000\) allow the agent to adapt quickly to the most recent workload trends. This combination results in a responsive agent that aggressively provisions resources to optimize for future Quality of Service, setting it apart from the more conservative, reactive baselines.

% \begin{figure}[tbp]
%     \centering
%     \includegraphics[width=\columnwidth]{assets/hpa_baseline.png} 
%     \caption{HPA (Reactive Baseline)}
%     \label{fig:hpa}
% \end{figure}

% \begin{figure}[tbp]
%     \centering
%     \includegraphics[width=\columnwidth]{assets/keda_baseline.png}
%     \caption{KEDA (Flexible Trigger)}
%     \label{fig:keda}
% \end{figure}

% \begin{figure}[tbp]
%     \centering
%     \includegraphics[width=\columnwidth]{assets/dqn_baseline_test.png}
%     \caption{NimbusGuard (Proactive)}
%     \label{fig:dqn}
% \end{figure}

As shown in Table~\ref{tab:comparision}, NimbusGuard operated with the highest average replica count (5.44), significantly more than HPA (3.05) and KEDA (2.93). This led to it having the largest resource integral (2,775 pod-seconds), indicating a strategy that prioritizes Quality of Service and responsiveness over minimizing cost.
Furthermore, NimbusGuard was the most agile and least stable system, executing 8 total scaling events, double that of HPA and KEDA (4 events each). In contrast, HPA and KEDA offered greater stability and resource efficiency, making them more cost-effective but potentially less responsive to sudden load spikes. These findings highlight a fundamental trade-off: the proactive, performance-focused scaling of NimbusGuard versus the reactive, cost-efficient stability of traditional autoscalers.

\begin{figure}[H]
    \centering
    \includegraphics[width=\columnwidth]{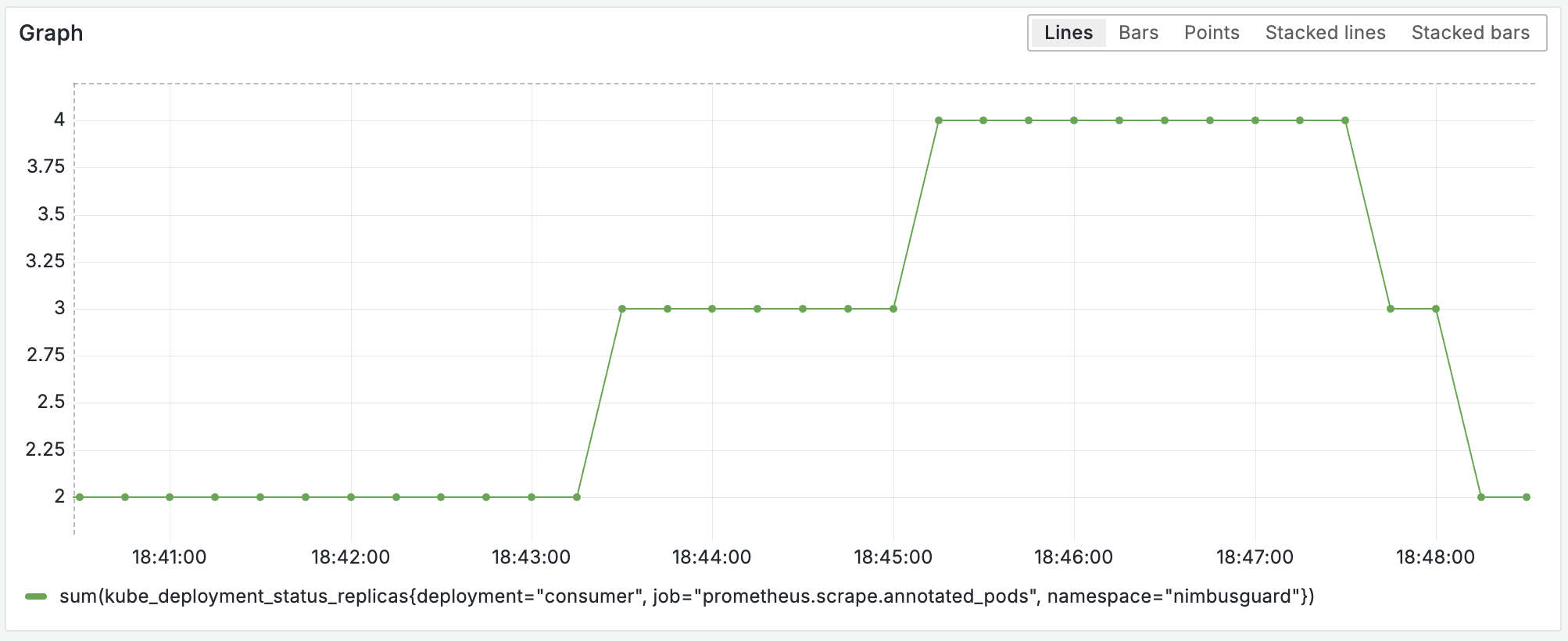} 
    \caption{HPA (Reactive Baseline)}
    \label{fig:hpa}
\end{figure}

\begin{figure}[H]
    \centering
    \includegraphics[width=\columnwidth]{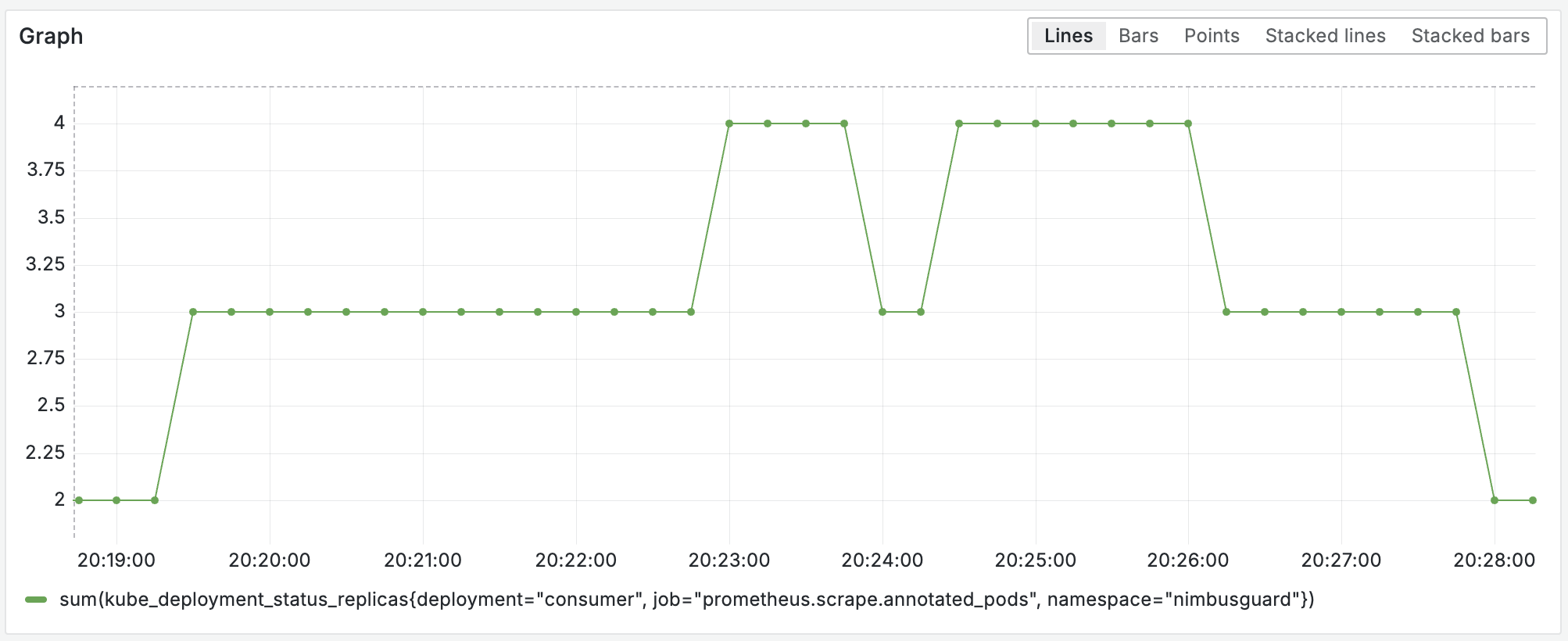}
    \caption{KEDA (Flexible Trigger)}
    \label{fig:keda}
\end{figure}

\begin{figure}[H]
    \centering
    \includegraphics[width=\columnwidth]{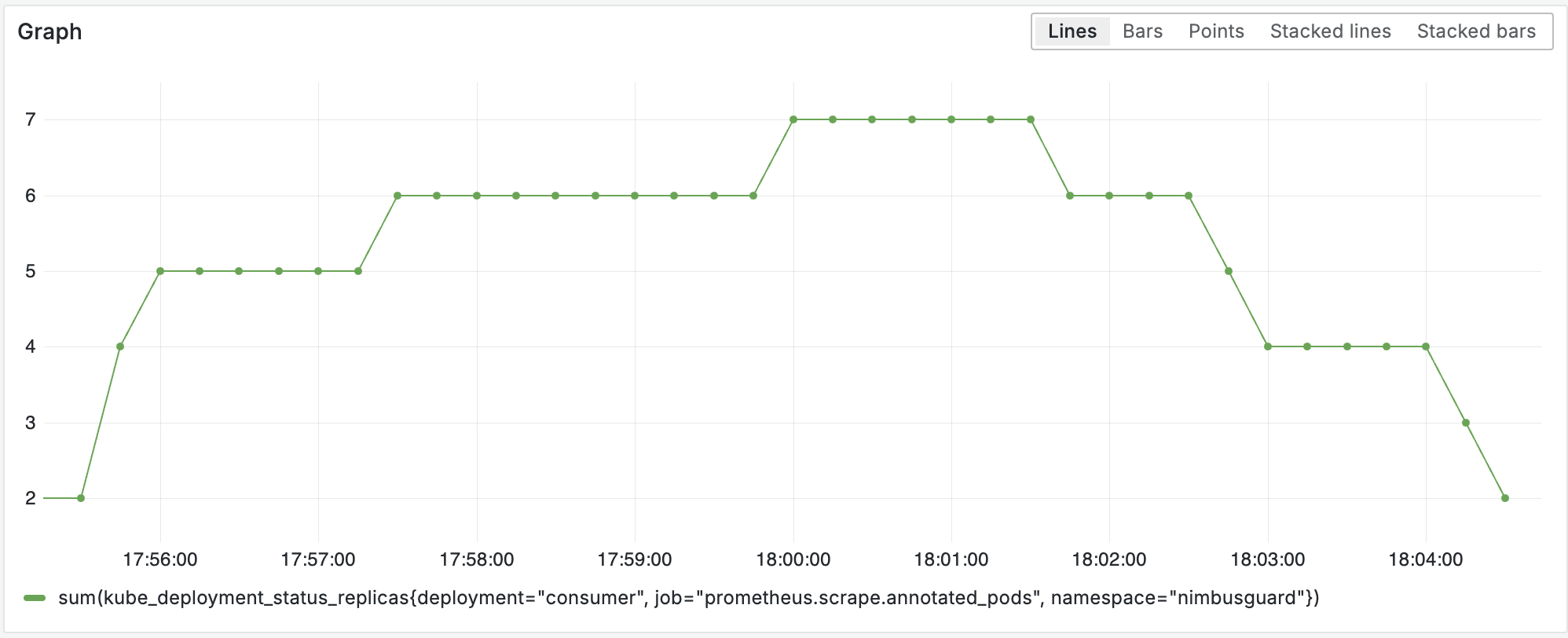}
    \caption{NimbusGuard (Proactive)}
    \label{fig:dqn}
\end{figure}

\begin{table}[htbp]
    \centering
    \caption{Experimental Performance Comparison}
    \label{tab:comparision}
    \resizebox{\columnwidth}{!}{%
    \begin{tabular}{@{}lccc@{}}
        \toprule
        \textbf{Performance Metric} & \textbf{DQN} & \textbf{HPA} & \textbf{KEDA} \\
        \midrule
        Avg. Time to Scale (sec)         & $\sim 60\,\mathrm{s} \pm 5\,\mathrm{s}$ & $\sim 300\,\mathrm{s} \pm 5\,\mathrm{s}$ & $\sim 90\,\mathrm{s} \pm 5\,\mathrm{s}$ \\
        Avg. Replicas (pods)     & 5.44  & 3.05  & 2.93  \\
        Peak Replicas (pods)        & 7     & 4     & 4     \\
        Total Scaling Events        & 8     & 4     & 4     \\
        \bottomrule
    \end{tabular}%
    }
    \footnotesize
    \vspace{0.1cm}
    \begin{flushleft}
    \textit{DQN represents the proposed NimbusGuard system with Deep Q-Network intelligence with an uncertainty of $\pm 5\,\mathrm{s}$}
    \end{flushleft}
\end{table}

\subsection{DQN-Specific Intelligence Analysis}

A deeper analysis of the agent's internal state provides insight into its decision-making process. The efficacy of the primary proactive scaling feature, predicted memory utilization ($s_1$), is fundamentally dependent on the accuracy of the underlying forecasting model. An enhanced LSTM model was developed to predict memory usage, a key component of the state vector. The model was trained on historical data from the target application and evaluated on a hold-out test set.

\begin{figure}[htbp]
\centering
% The image width is set to 80% of the column's width.
\includegraphics[width=1\columnwidth]{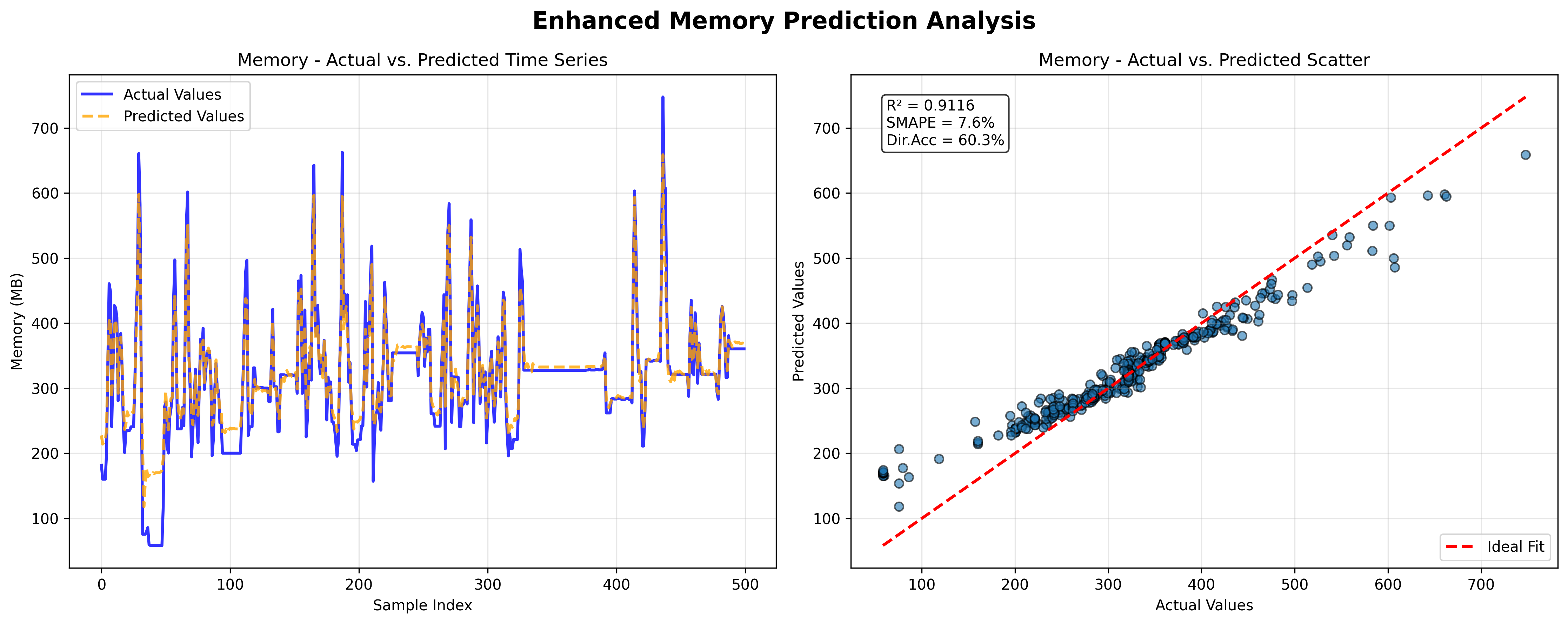}
\caption{LSTM Feature Analysis shows the proactive forecasts for load pressure and trend velocity.}
\label{lstm_forecast}
\end{figure}

Fig. \ref{lstm_forecast} Analysis of the enhanced memory prediction model. The time series plot (left) shows a close tracking between actual and predicted values. The scatter plot (right) shows data points clustering tightly around the ideal fit line, visually confirming the high R² value. 

\begin{figure}[htbp]
\centering
% The image width is set to 80% of the column's width.
\includegraphics[width=1\columnwidth]{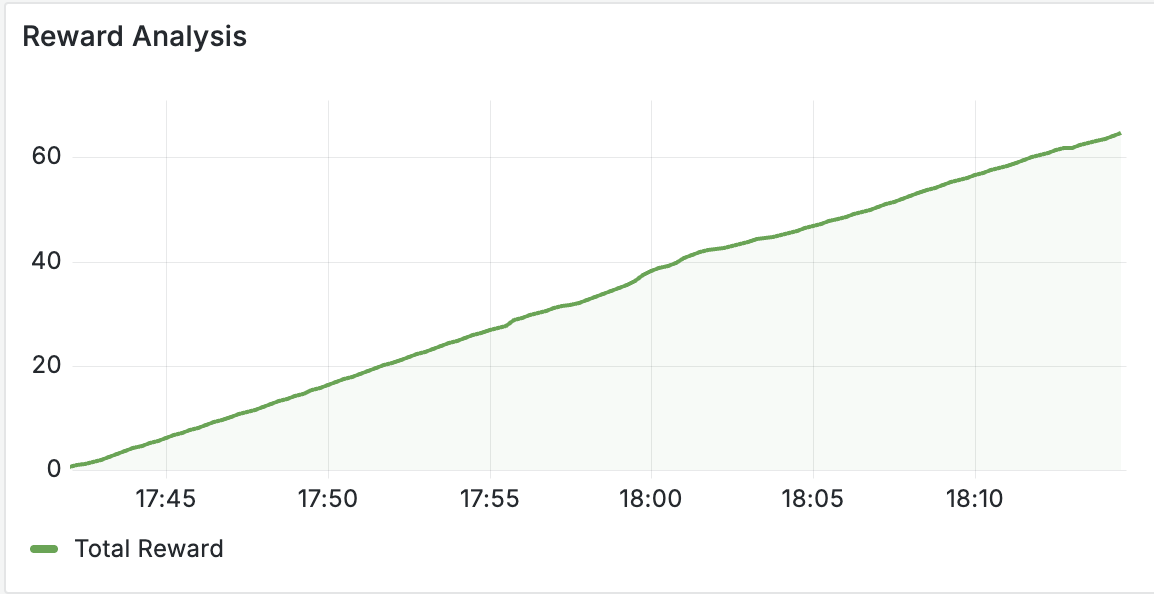}
\caption{Reward Analysis shows the evolution of the reward signal, which guides the agent's learning.}
\label{reward_signal}
\end{figure}

Fig. \ref{reward_signal} shows the immediate reward the agent received after each decision. The signal is a composite score derived from the multi-objective reward function, balancing performance, cost, and stability. The fluctuations reflect the constant trade-offs the agent must make. For example, a negative dip might correspond to a moment of temporary pod unavailability during a scale-up, which penalizes the agent and teaches it to scale more carefully in the future.

% \begin{figure}[htbp]
% % Use \centering for better spacing control within the figure environment.
% \centering
% % Set the width of the image to the width of the current column.
% \includegraphics[width=\columnwidth]{../assets/reward_analysis.png}
% \caption{Reward Analysis shows the evolution of the reward signal, which guides the agent's learning.}
% \label{reward_signal}
% \end{figure}

The LSTM forecasts (Fig. \ref{lstm_forecast}) provided the agent with proactive signals about future load pressure, enabling it to prepare for changes rather than just react to them. The divergence of these lines shows the agent learning to prefer actions that it predicts will lead to higher future rewards. The reward signal (Fig. \ref{reward_signal}) itself, while fluctuating, demonstrates the feedback loop that drives this learning process.
% \section{Discussion}
% The results demonstrate the algorithmic advantages of NimbusGuard's design. The multi-objective reward function successfully balanced competing goals, while the proactive LSTM forecasting and clean signals from the asynchronous load generation allowed the system to anticipate load changes. This capability is inherently lacking in reactive systems like HPA and KEDA.

% While the study was conducted on a multi-node cluster, a limitation is the sequential evaluation of the autoscalers. Future work should aim for simultaneous A/B testing to eliminate any potential for temporal bias and to validate performance under more complex, interfering network conditions. Furthermore, several avenues for algorithmic enhancement exist. The current action space is limited to scaling by a single replica (+1, 0, -1). Future research could explore an expanded action space (e.g., +2, -2) to allow the agent to react more aggressively to sudden, large load spikes, potentially improving responsiveness. Additionally, this work focuses exclusively on horizontal scaling (changing the number of pods). A promising direction for future work is to integrate Vertical Pod Autoscaling (VPA), creating a hybrid system where the agent can learn to decide whether it is more efficient to add more pods (horizontal) or to increase the resources (CPU/memory) of existing pods (vertical).

\section{Discussion}
The results confirm NimbusGuard's algorithmic advantages; its multi-objective reward function and proactive LSTM forecasting enable it to anticipate load changes, a capability reactive systems like HPA and KEDA lack. However, the study is limited by the sequential evaluation of the autoscalers.

Future work should focus on two key areas:
\begin{itemize}
    % \item Conducting simultaneous A/B testing to eliminate temporal bias.
    \item Expanding the action space (e.g., +2/-2 replicas) to allow for more aggressive responses to load spikes.
    \item Integrating Vertical Pod Autoscaling (VPA) to create a hybrid system that can choose between adding more pods or increasing the resources of existing ones.
\end{itemize}
% While this proactive approach resulted in an exceptionally fast reaction time of just 60 seconds, it came at the cost of a 62\% increase in resource consumption compared to HPA and double the number of scaling events, indicating lower stability. These findings demonstrate the potential of deep reinforcement learning to create autonomous, policy-aware systems that can be customized for specific business goals, such as maximizing performance or minimizing costs.

% The DQN-based autoscaler, NimbusGuard, successfully implemented a novel scaling strategy that prioritizes performance and responsiveness over resource efficiency. While this proactive approach resulted in an exceptionally fast reaction time of just 60 seconds, it came at the cost of a 62\% increase in resource consumption compared to HPA and double the number of scaling events, indicating lower stability. These findings demonstrate the potential of deep reinforcement learning to create autonomous, policy-aware systems that can be customized for specific business goals, such as maximizing performance or minimizing costs.
% \subfile{sections/conclusion}

\bibliographystyle{IEEEtran}
\bibliography{sections/references}

\begin{thebibliography}{10}
\providecommand{\url}[1]{#1}
\csname url@samestyle\endcsname
\providecommand{\newblock}{\relax}
\providecommand{\bibinfo}[2]{#2}
\providecommand{\BIBentrySTDinterwordspacing}{\spaceskip=0pt\relax}
\providecommand{\BIBentryALTinterwordstretchfactor}{4}
\providecommand{\BIBentryALTinterwordspacing}{\spaceskip=\fontdimen2\font plus
\BIBentryALTinterwordstretchfactor\fontdimen3\font minus
  \fontdimen4\font\relax}
\providecommand{\BIBforeignlanguage}[2]{{%
\expandafter\ifx\csname l@#1\endcsname\relax
\typeout{** WARNING: IEEEtran.bst: No hyphenation pattern has been}%
\typeout{** loaded for the language `#1'. Using the pattern for}%
\typeout{** the default language instead.}%
\else
\language=\csname l@#1\endcsname
\fi
#2}}
\providecommand{\BIBdecl}{\relax}
\BIBdecl

\bibitem{ind1}
N.~Kratzke and P.-C. Quint, ``Understanding cloud-native applications after 10
  years of cloud computing-a systematic mapping study,'' \emph{Journal of
  Systems and Software}, vol. 126, pp. 1--16, 2017.

\bibitem{ind5}
M.~A. Tamiru, J.~Tordsson, E.~Elmroth, and G.~Pierre, ``An experimental
  evaluation of the kubernetes cluster autoscaler in the cloud,'' in \emph{2020
  IEEE International Conference on Cloud Computing Technology and Science
  (CloudCom)}.\hskip 1em plus 0.5em minus 0.4em\relax IEEE, 2020, pp. 17--24.

\bibitem{cham10}
``\BIBforeignlanguage{en}{Kubernetes},'' \url{https://kubernetes.io/}, 2025,
  [Accessed 2025-07-28].

\bibitem{ind2}
E.~A. Brewer, ``Kubernetes and the path to cloud native,'' in \emph{Proceedings
  of the sixth ACM symposium on cloud computing}, 2015, pp. 167--167.

\bibitem{ind3}
R.~Aurangzaib, W.~Iqbal, M.~Abdullah, F.~Bukhari, F.~Ullah, and A.~Erradi,
  ``Scalable containerized pipeline for real-time big data analytics,'' in
  \emph{2022 IEEE International Conference on Cloud Computing Technology and
  Science (CloudCom)}.\hskip 1em plus 0.5em minus 0.4em\relax IEEE, 2022, pp.
  25--32.

\bibitem{ind4}
L.~Toka, G.~Dobreff, B.~Fodor, and B.~Sonkoly, ``Machine learning-based scaling
  management for kubernetes edge clusters,'' \emph{IEEE Transactions on Network
  and Service Management}, vol.~18, no.~1, pp. 958--972, 2021.

\bibitem{cham9}
``{L}ang{G}raph --- langchain.com,'' \url{https://www.langchain.com/langgraph},
  [Accessed 30-07-2025].

\bibitem{ind6}
``{G}it{H}ub - kedacore/keda: {K}{E}{D}{A} is a {K}ubernetes-based {E}vent
  {D}riven {A}utoscaling component. {I}t provides event driven scale for any
  container running in {K}ubernetes --- github.com,''
  \url{https://github.com/kedacore/keda}, [Accessed 28-07-2025].

\bibitem{ind7}
``{K}{E}{D}{A} --- cncf.io,'' \url{https://www.cncf.io/projects/keda/},
  [Accessed 28-07-2025].

\bibitem{ind8}
F.~Rossi, V.~Cardellini, and F.~L. Presti, ``Hierarchical scaling of
  microservices in kubernetes,'' in \emph{2020 IEEE international conference on
  autonomic computing and self-organizing systems (ACSOS)}.\hskip 1em plus
  0.5em minus 0.4em\relax IEEE, 2020, pp. 28--37.

\bibitem{cham1}
S.~K. Mondal, X.~Wu, H.~M.~D. Kabir, H.-N. Dai, K.~Ni, H.~Yuan, and T.~Wang,
  ``Toward optimal load prediction and customizable autoscaling scheme for
  kubernetes,'' \emph{Mathematics}, vol.~11, no.~12, p. 2675, 2023.

\bibitem{cham2}
N.-M. Dang-Quang and M.~Yoo, ``Deep learning-based autoscaling using
  bidirectional long short-term memory for kubernetes,'' \emph{Applied
  Sciences}, vol.~11, no.~9, p. 3835, 2021.

\bibitem{cham3}
A.~A. Khaleq and I.~Ra, ``Intelligent autoscaling of microservices in the cloud
  for real-time applications,'' \emph{IEEE access}, vol.~9, pp.
  35\,464--35\,476, 2021.

\bibitem{ind9}
M.~Imdoukh, I.~Ahmad, and M.~G. Alfailakawi, ``Machine learning-based
  auto-scaling for containerized applications,'' \emph{Neural Computing and
  Applications}, vol.~32, no.~13, pp. 9745--9760, 2020.

\bibitem{ind10}
M.~Yan, X.~Liang, Z.~Lu, J.~Wu, and W.~Zhang, ``Hansel: Adaptive horizontal
  scaling of microservices using bi-lstm,'' \emph{Applied Soft Computing}, vol.
  105, p. 107216, 2021.

\bibitem{cham6}
D.-D. Vu, M.-N. Tran, and Y.~Kim, ``Predictive hybrid autoscaling for
  containerized applications,'' \emph{IEEE Access}, vol.~10, pp.
  109\,768--109\,778, 2022.

\bibitem{cham4}
Y.~Gar{\'\i}, D.~A. Monge, and C.~Mateos, ``A q-learning approach for the
  autoscaling of scientific workflows in the cloud,'' \emph{Future Generation
  Computer Systems}, vol. 127, pp. 168--180, 2022.

\bibitem{cham5}
Z.~Jian, X.~Xie, Y.~Fang, Y.~Jiang, Y.~Lu, A.~Dash, T.~Li, and G.~Wang, ``Drs:
  A deep reinforcement learning enhanced kubernetes scheduler for
  microservice-based system,'' \emph{Software: Practice and Experience},
  vol.~54, no.~10, pp. 2102--2126, 2024.

\bibitem{cham7}
G.~Zhang, W.~Guo, Z.~Tan, Q.~Guan, and H.~Jiang, ``Kis-s: A gpu-aware
  kubernetes inference simulator with rl-based auto-scaling,'' \emph{arXiv
  preprint arXiv:2507.07932}, 2025.

\bibitem{ind11}
S.~Subramaniam and G.~H. Loh, ``Fire-and-forget: Load/store scheduling with no
  store queue at all,'' in \emph{2006 39th Annual IEEE/ACM International
  Symposium on Microarchitecture (MICRO'06)}.\hskip 1em plus 0.5em minus
  0.4em\relax IEEE, 2006, pp. 273--284.

\bibitem{cham8}
``asyncio {A}synchronous {I}/{O},''
  \url{https://docs.python.org/3/library/asyncio.html}, [Accessed 30-07-2025].

\end{thebibliography}

\end{document}